\documentclass{kapproc} % Computer Modern font calls

\usepackage{procps} 

\usepackage[dvips]{graphicx}

\upperandlowercase

 \let\footnote\savefootnote

\kluwerbib % will produce this kind of bibliography entry:

\begin{document}

\articletitle[Disk galaxy evolution]{Eight billion years of disk galaxy \\
evolution}

\articlesubtitle{No galaxy is an island}

\author{Eric F.\ Bell,\altaffilmark{1} Marco Barden,\altaffilmark{1},
Xianzhong Zheng\altaffilmark{1}, Casey Papovich\altaffilmark{2},
Emeric Le Floc'h\altaffilmark{2}
George Rieke\altaffilmark{2}, Christian Wolf\altaffilmark{3}, and the GEMS, MIPS Instrument,
and COMBO-17 teams}

\affil{\altaffilmark{1}Max-Planck-Institut f\"ur Astronomie, Heidelberg, Germany {\rm bell@mpia.de} \ 
\altaffilmark{2}Steward Observatory, University of Arizona, Tuscon, AZ, USA \
\altaffilmark{3}University of Oxford, Oxford, UK }

\begin{abstract}
We present a brief discussion of the evolution of disk
galaxy stellar masses, sizes, rotation velocities, and 
star formation rates over the last eight billion years.
Recent observations have failed to detect significant evolution in the 
stellar mass Tully-Fisher
relation, stellar mass--size relation, and the 
stellar mass function of disk galaxies.  Yet, most
$z<1$ star formation is in disks, and this star formation 
would be expected to drive a rapid growth of the total 
stellar mass (and therefore mass function) of disks in the 
last eight billion years.  Such a build-up is not seen; instead, 
a rapid build-up in the total stellar mass in non-star-forming
spheroid-dominated galaxies is observed.  
Large numbers of disk-dominated galaxies are systematically 
shutting off their star formation and building up a spheroid (or losing 
a disk) in the epoch $0<z<1$.
\end{abstract}


The evolution of disk galaxy scaling relations --- e.g., 
the luminosity--line\-width (Tully-Fisher) relation, or
the luminosity--size or stellar mass--size relations ---
give insight into the evolution of the masses, stellar
populations and angular momentum contents of galaxy disks.
Luminosity and stellar mass functions give yet deeper insight, 
as they show how galaxies populate
these scaling relations.  
Here, we present a very brief
overview of some recent insights into the evolution of disk 
galaxy scaling relations, their stellar mass function, 
and to explore the interplay between star formation and the growth
of stellar mass.  We will consider only $z<1$, or the last
eight billion years or so, given a concordance
cosmology ($\Omega_{m,0} = 0.3$, $\Omega_{\Lambda,0} = 0.7$, and 
      ${\rm H_0 = 70\,km\,s^{-1}\,Mpc^{-1}}$).

\section{(Non)-evolution of disk galaxy scaling relations}

The evolution of disk galaxy scaling relations has proven
difficult to robustly determine, and some issues remain 
contentious.  In some cases (the Tully-Fisher relation: 
Vogt et al.\ 1996 vs.\ e.g., 
Boehm et al.\ 2004) it is not clear where the source of 
the discrepancies lie.  In other cases (the luminosity/stellar mass--size
relations: e.g., Ravindranath et al.\ 2004 vs. Barden et al.\ 2005) the 
source of the discrepant interpretations has been resolved (see 
the discussion in Barden et al.\ 2005). 

\cite{barden05} present an analysis of the evolution of the 
disk galaxy luminosity--size, and stellar mass--size relation
over the epoch $0<z<1$ from the HST/GEMS (Galaxy 
Evolution from Morphology and SEDs; Rix et al.\ 2004) survey
of the Extended Chandra Deep Field South.  
They chart the evolution of the distribution of disk sizes
from $z=0$ (using the SDSS) to gradually higher and higher
redshift, carefully quantifying the effects of surface brightness
dimming and incompleteness.  At $z>1.1$, 
the surface brightness limit of GEMS starts to significantly 
eat into the disk galaxy population, therefore they limited
their study to $z<1.1$.  Using this sample of galaxies with 
well-understood and modest completeness corrections, they conclude that:
\begin{itemize}
\item there has been significant evolution in 
the luminosity--size relation, such that galaxies at
$z=1$ have roughly 1.5\,mag\,arcsec$^{-2}$ brighter
rest-frame $B$-band surface brightnesses than galaxies
of equivalent luminosity today; and
\item that most of this evolution is attributable 
to the fading and reddening of stellar populations
as they age from $z=1$ to the present day, i.e., 
there is no detectable evolution in the stellar mass--size
relation for galaxies with $M_* > 10^{10} M_{\odot}$.  
\end{itemize}

The evolution of the Tully--Fisher relation 
remains a contentious issue.  None\-the\-less, important
progress is still being made.  Novel contributions 
were made recently by \cite{conselice05} and \cite{flores06}, 
who presented studies of the $z<1$ evolution of the 
near-IR and stellar mass Tully-Fisher relations.  
Their analyses ruled out substantial
evolution in the stellar mass Tully-Fisher relation\footnote{Conselice 
et al.\ then use this as an argument that the dark matter content
and baryonic content of disks grow in lockstep, arguing that 
the stars were a good probe of baryonic mass and the rotation 
velocity was a good probe of dark mass.  Yet, it is important to 
note that if the bulk of massive disks are maximum-disk (i.e., 
their baryonic content dominates their rotation velocity; e.g.,
Kassin et al.\ 2006, Kassin et al.\ this volume) then one also expects
no evolution in the stellar mass Tully-Fisher relation.}.

Also recently, a number of groups have studied the
evolution of the stellar mass function of galaxies
split either by morphological type or color.  These
cuts are, at a very general level, largely equivalent ---
most red galaxies are morphologically early-type
and the bulk of blue galaxies are morphologically
late-type out to $z=0.7$ at least (Strateva et al.\ 2001; Bell et al.\ 2004).
A general conclusion is that the stellar mass function of 
blue/disk galaxies does not evolve significantly
since $z=1$ (see Brinchmann \& Ellis 2000 for 
first indications of this result; Bundy et al.\ 2006; 
Borch et al.\ 2006)\footnote{The oft-cited $\sim 1.5$\,magnitude
fading of the characteristic luminosity $L^*$ of the 
disk galaxy population from $z=1$ to the present
is more-or-less accounted for by the expected fading and reddening
of the ageing stellar populations in blue star-forming disks:
thus, the characteristic stellar mass of disks does not 
appear to evolve a large amount during the last 8 billion years
(Lilly et al.\ 1995; Willmer et al.\ 2006; Borch et al.\ 2006; 
Blanton 2006)}.

It would appear that none of the scaling relations
defining disk galaxy structure or dynamics, or indeed
the space density of disk galaxies at a given mass, 
significantly change in the last eight billion years.  
{\it It is as if the disk galaxy population
resolved to do nothing for the last eight billion years
except age.}

Yet, there is one observational constraint that
has not yet been brought to bear on the problem, which 
seems to give interesting and incisive insight: the 
evolution of disk galaxy star formation rates.

\section{Eight billion years of star formation in disks}

With the advent of Spitzer, it has been possible
to explore the star formation rates of galaxies
at $z<1$ with unprecedented accuracy: fully 50\%
of the $z=1$ cosmic SFR has been resolved into 
individual galaxies with deep 24$\mu$m surveys
(Le Floc'h et al.\ 2005; Zheng et al.\ 2006).  
One key result is that 
the bulk of $z<1$ star formation is in disk galaxies:
{\it the order-of-magnitude decline in cosmic
SFR is the result of processes which are shaping the evolution 
of disk galaxies} (Bell et al.\ 2005; Melbourne et al.\ 2005;
Wolf et al.\ 2005).

We explore this issue further in Fig.\ 1.  Here, 
we show the contribution of different types of galaxy
to the cosmic SFR and stellar mass budget at $z<1$: such an
analysis requires the assumption of a universally-applicable 
stellar IMF, and we adopt the parameterization of 
Chabrier (2003) in what follows. 
Full circles show the total SFR/stellar mass.  Open
diamonds show the estimated contribution from early-type
galaxies.  These were selected to be on the red sequence and 
have concentrated light profiles, with Sersic indices $n>2.5$ 
derived from HST/ACS F850LP imaging data from GEMS.  This selection 
is analogous to that used by McIntosh et al.\ (2005) for their 
study of the early-type galaxy stellar mass--size relation.
Asterisks show the contribution all other galaxies, dominated by 
$n<2.5$ galaxies (i.e., 
less concentrated, typically disk-dominated galaxies) with 
a small contribution from blue galaxies with $n>2.5$ (which 
are often blue disks with significant bulges, or ongoing galaxy 
mergers). Other ways of splitting the sample --- e.g., 
a pure $n=2.5$ split, a pure blue/red split, or splits by visual 
morphology --- produce qualitatively similar results to those
shown in Fig.\ 1.  The total stellar mass budget is
derived from three COMBO-17 photometric redshift
survey fields (see Borch et al.\ 2006); the type-split stellar mass
and all SFR budgets used in the 
argument below were derived from only one field, the 
extended Chandra Deep Field South for which both HST/ACS
data and Spitzer data were available.  Uncertainties
from cosmic variance have been estimated by exploring field-to-field 
variation in stellar mass budget of the red and blue galaxy populations
(from COMBO-17).  The $z=0$ point is taken from the SDSS/2MASS 
(Bell et al.\ 2003).

In the upper panel of 
Fig.\ 1 we show the evolution of the cosmic star
formation rate for all galaxies derived from 
Spitzer 24$\mu$m observations (solid circles; 
Le Floc'h et al.\ 2005; Bell et al.\ in prep.; and 
grey error bars show the compilation from 
Hopkins 2004).  The asterisks show the evolution of the 
integrated SFR density in blue/disk-dominated galaxies; the 
diamonds the contribution from red, early-type galaxies.  
It is clear that
the bulk of star formation is in blue and disk galaxies at all $z<1$.

What do these SFRs predict for the build-up of stellar
mass in disks?  This is explored in the lower panels
of Fig.\ 1, where we compare the evolution of the
integrated stellar mass density with the integral of the star
formation rate of different galaxy populations.  
The evolution of the integrated stellar
mass density in all galaxies (solid circles), 
blue and disk galaxies (asterisks) and red 
spheroid-dominated galaxies (diamonds)
is shown.  The lines show the integral of the cosmic SFR:
clearly, the evolution of the total cosmic SFR (solid line) is compatible
with the observed build-up in cosmic stellar mass density
from $z=1$ to the present day, with some 40\% of stars being 
formed in this epoch.

Disk galaxies contain the bulk of the 
star formation.  Thus if one assumes that the disk 
galaxy population is a closed box (i.e., all stars
ever formed in disks stay there) then one can 
predict the growth in the total disk stellar mass density.
This result is shown by the dotted line.  It is difficult
to argue that this mismatch is some kind of catastrophic
failure of the stellar mass/SFR estimation methodology: 
completely standard calibrations were used, and the integral
of the cosmic SFR indeed reproduces well the build-up of 
stellar mass at $z<1$.  Instead, it
is clear that {\it the disk galaxy population is
not a closed box.}  The key to understanding this issue
lies in exploring the stellar mass growth in spheroid-dominated
galaxies: the total stellar mass in spheroids increases
significantly from $z=1$ to the present day, while almost no star formation 
happens in the spheroid-dominated galaxy population.  
{\bf Large numbers of disk-dominated galaxies are systematically 
shutting off their star formation and building up a spheroid (or losing 
a disk)} in the epoch $0<z<1$.  The physical processes that are 
responsible for these transformations are as yet unclear, and 
might include bar-induced 
instabilities, merging with satellites or companions, or 
suppression of star formation by ram-pressure stripping or galactic winds.

\begin{figure}[ht]
\centerline{\includegraphics[width=3.0in]{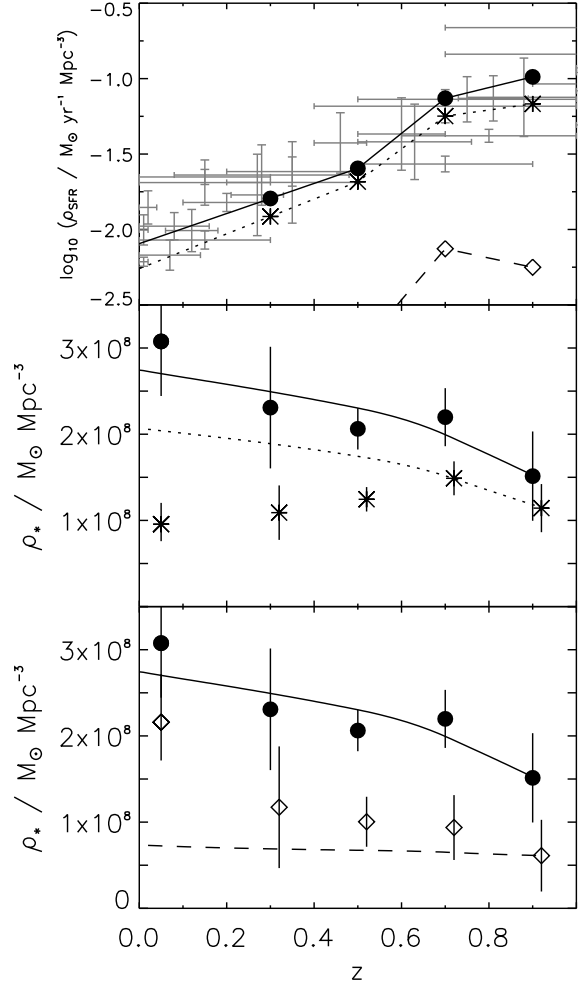}}
\vspace{-1.4cm}
\caption{ The cosmic evolution of 
star formation rate and stellar mass (filled circles and solid
lines), split into 
contributions from the red, morphologically early-type 
galaxies (diamonds and dashed lines) and all other galaxies 
(asterisks and dotted lines; dominated by disk galaxies).  
In the uppermost
panel, determinations of cosmic SFR are shown also in grey, 
adopted from Hopkins (2004).  In the lower panels, the data points
show the observed evolution of stellar mass as a function of redshift.
The solid line shows the predicted build-up of stellar mass, assuming
the star formation history shown by the solid line in the upper panel 
(assuming for gas recycling); it is clear that the integral of the cosmic
star formation history predicts the cosmic evolution of stellar 
mass at $z<1$ reasonably accurately.  The dotted line 
shows the predicted
evolution of total stellar mass assuming that all stars that form 
in blue/disk galaxies remain in blue/disk galaxies (i.e., if the blue cloud
evolves like a closed box).  The dashed line shows the corresponding 
evolution for early-type galaxies.}
\end{figure}

\section{Discussion}

This has important implications for how one interprets
the non-evolution of disk galaxy scaling relations.
Whereas one could have, without the measurement of the 
disk galaxy SFRs, postulated that disks were a 
non- or slowly-evolving population, the evolution 
of such (expected) rapid star formation-driven build-up in mass
makes this interpretation meaningless.
Instead, as low-mass disk galaxies
are growing in mass (as they must, to replace 
the disks sacrificed for the ever-growing 
spheroid-dominated galaxy population), they must 
be growing also in rotation velocity (to preserve
the non-evolving Tully-Fisher relation) and size 
(to preserve the non-evolving mass--size relation) --- at 
least in an average sense.  Furthermore, in this picture, 
the non-evolution of the stellar mass function of disks
is a vital constraint: any mechanism which is transforming
disks into spheroids must `take' disks from the stellar mass
functions at such a rate that they can be approximately
replaced by star formation-related growth of lower-mass disks.

\end{document}